\documentstyle[12pt,epsf]{article}
\def\clock{\count0=\time \divide\count0 by 60
     \count1=\count0 \multiply\count1 by -60 \advance\count1 by \time
     \number\count0:\ifnum\count1<10{0\number\count1}\else\number\count1\fi}

\begin{document}
\newcommand{\rg}{\sqrt{g}}
\newcommand{\etal}{{\it et al }}
\newcommand{\Nabla}{\bigtriangledown}
\def\lesssim{\mathrel{\hbox{\rlap{\hbox{\lower4pt\hbox{$\sim$}}}\hbox{$<$}}}}
\def\gtrsim{\mathrel{\hbox{\rlap{\hbox{\lower4pt\hbox{$\sim$}}}\hbox{$>$}}}}
\let\la=\lesssim
\let\ga=\gtrsim

\title{Topological Defects with Broken Scale Invariance}
\author{Ue-Li Pen}
\date{}
\maketitle

\begin{center}
{\it Harvard-Smithsonian Center for Astrophysics}
\end{center}
\vspace{0.3in}

\begin{abstract}
Defect models have recently been declared dead\cite{watson97}, because
they predict microwave background and matter fluctuations grossly out
of line with what we see. In this talk we apply the fact that
many defects are automatically destroyed at the time of
radiation-matter transition, thus resurrecting the defects model.
Moreover, the resurrected version predicts a cosmological constant,
explains the apparent excess of hot clusters and the non-Gaussianity
observed in galaxy surveys.  If this model is correct, then the MAP
and PLANCK missions will not measure what people expect them to
(oscillations); rather, they will measure a broad hump.
\end{abstract}

The traditional class of scale invariant defect models has severe
problems\cite{pst97}.  They do not explain the rise in fluctuation
amplitude at scales of a few degrees, and the slope of the
perturbations is at conflict with the observed matter power spectrum.
We show here that global topological defects are generically destroyed
at the matter-radiation transition at a cosmic age of 100000 years.
In this case, instead of being at great discrepancy with the cosmic
microwave background radiation fluctuations, this model in fact
correctly explains several hitherto mysterious phenomena.  The most
recent supernovae search\cite{science,perlmutter}, concordance
arguments\cite{oststei} and direct angular diameter distance
measurements\cite{danos} have suggested that the universe is filled
with a cosmological constant.  Its existence is at best mysterious in
most cosmological models.  The new defect models, however, predict
such a cosmological constant from the COBE observations.

Let us now address the decay of global defects at matter-radiation
equality.  In modern particle physics, all interactions which are not
forbidden actually occur.  On this ground, one generically expects
couplings between gravity and other quantum fields which are stronger
than just the space time curvature.  Global fields experience
couplings of the form $\epsilon\phi_1^\dagger\phi_2 R$, where
$\epsilon$ is a dimensionless coupling constant of order unity, and
$R$ is the scalar curvature.  In the radiation epoch of the universe,
$R= 0$, so the coupling has no dynamical effect.  But when the
universe enters matter domination, this coupling will generally
destroy the exact symmetry which gave rise to the defects, and they
will collapse coherently.  During the onset of collapse, the vacuum
energy contributed by the new coupling will boost the power on the
largest scales, making the power spectrum more consistent with the
observed excess of large scale power.  Today there would be no defects
remaining.  At late times when the cosmological constant begins to
dominate, fluctuations decay, thus inducing large scale CMB
anisotropies.  The COBE detection in this model would be solely due to
relatively local, sub-horizon, fluctuations at the ca 100 Mpc scale,
whose potential decays at late time through the influence of a
cosmological constant.  Since linear perturbations do not generate any
late time anisotropies in a flat matter dominated universe, the COBE
detection in fact predicts a cosmological constant.  This is in
contrast to inflationary models, where a cosmological constant is
disfavoured by the COBE observations \cite{bunn}.

These defect models have two free parameters.  One is the symmetry
breaking scale $\phi_0$, and the second is its coupling to the Riemann
scalar $\epsilon$.  From the velocity dispersion in clusters of
galaxies we expect $\phi_0/m_{\rm planck} \sim v^2/c^2 \sim 10^{-5}$,
which would also be related to the COBE fluctuations.  This energy
scale suggests that the defect would be relicts of the breaking of the
Grand-Unified-Theory in the early universe.  A predictive property of
this model is the non-Gaussianity of fluctuations.  The strongest
constraints arises from the statistics of rich clusters\cite{chiu97}.
Several clusters with temperatures above 12 keV have been
found\cite{donahue}, and a recent cluster with a temperature of 17
keV\cite{tucker97} has been discovered.  The existence of these very
hot clusters is becoming an enigma for any Gaussian cosmological
model.  In non-Gaussian defect models, one would expect a larger
proportion of high temperature clusters relative to low temperature
ones.  The recent Las Campanas galaxy survey discovered features in
the two dimensional galaxy power spectrum which is inconsistent with
Gaussianity\cite{landy}.  Such signs of non-Gaussianity are
generically predicted in these defect models.

\begin{figure}
\epsfxsize=\hsize
\epsfbox{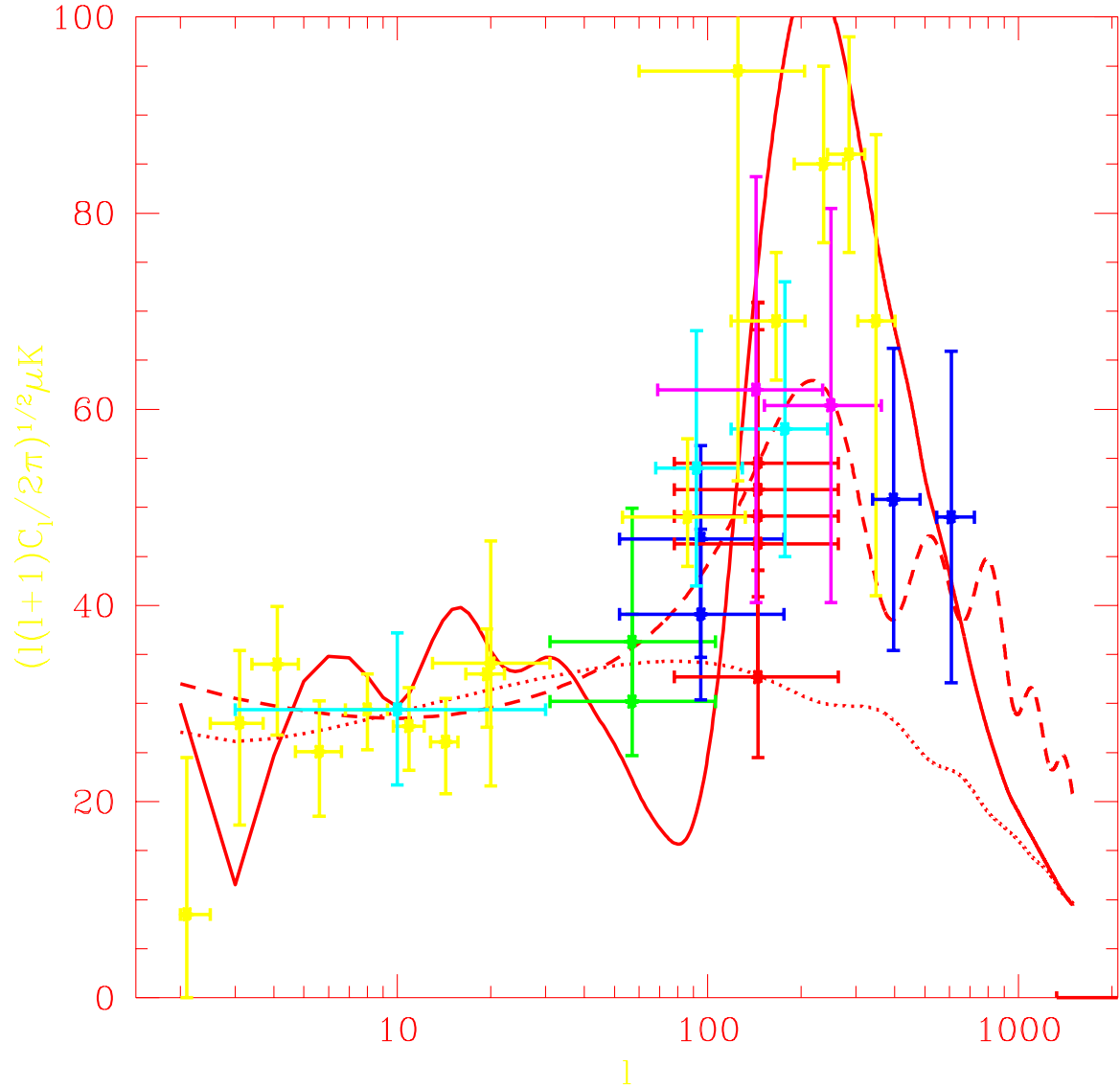}
\caption{
The angular CMB power spectrum.  The solid line shows the general
prediction of the broken global defect model.  It resolves the
problems exhibited by scaling defect models (dotted curve).  The
dashed line is the prediction of the cosmic concordance model of
Ostriker and Steinhardt\protect\cite{oststei}.  Overlaid on this plot
are the various experimental data points.}
\label{fig:cmb}
\end{figure}

In the defect scenario, the small scale CMB experiments MAP and Planck
will not measure the multiple oscillations expected from adiabatic
perturbations imprinted before the big bang, as predicted by
inflation.  Instead, the degree scale anisotropies are caused by the
incoherent properties of global defects\cite{pst97}.  A sample angular
power spectrum in such a scenario is shown in Figure \ref{fig:cmb}.
The effect of these broken defects is simulated using CMBFAST\cite{sz}
by truncating the primordial power at matter-radiation equality, which
corresponds to the peak in the power.  This prediction is heuristic,
since decoherence, vector modes and other complications are only
modelled in a simple fashion, but one would expect the qualitative
results to remain similar.  Without the oscillations, it will be very
difficult to reconstruct cosmological parameters accurately.  On the
other hand, this would be an exciting measurement of fundamental
physics and the history of symmetry breaking.

{\sc Acknowledgements}.  This work was supported by the Harvard Milton
fund.  The idea of broken scale invariance from non-minimal coupling is
due to Neil Turok.  I thank Uros Seljak for helpful discussions.

\end{document}